\documentclass[10pt,preprint]{aastex}
\def\re{{\rm Re}}
\def\im{{\rm Im}}
\def\gtwid{\mathrel{\raise.3ex\hbox{$>$\kern-.75em\lower1ex\hbox{$\sim$}}}}
\def\ltwid{\mathrel{\raise.3ex\hbox{$<$\kern-.75em\lower1ex\hbox{$\sim$}}}}
\def\haf{{\textstyle{{1}\over{2}}}}
\def\thd{{\textstyle{{1}\over{3}}}}
\def\qua{{\textstyle{{1}\over{4}}}}

\def\otf{{\textstyle{{1}\over{24}}}}

\def\pmb#1{\setbox0=\hbox{#1}%
  \kern-.025em\copy0\kern-\wd0
  \kern.05em\copy0\kern-\wd0
  \kern-.025em\raise.0433em\box0 }
\shortauthors{GWINN}
\shorttitle{CORRELATION STATISTICS}
\begin{document}
\input epsf

\title{CORRELATION STATISTICS OF QUANTIZED NOISELIKE SIGNALS}
\author{Carl Gwinn}
\affil{Department of Physics, University of California, Santa Barbara, 
California 93106, {\sf cgwinn@condor.physics.ucsb.edu} }

\vskip 1 truein
\begin{abstract}

I calculate the statistics of correlation of two digitized noiselike signals:
drawn from complex Gaussian distributions, sampled, quantized, 
correlated, and averaged.  
Averaged over many such samples, the correlation $r$ approaches a
Gaussian distribution. The mean and variance of $r$ fully characterize
the distribution of $r$.
The mean corresponds to the reproducible part of the
measurement, and the variance corresponds to the random
part, or noise. 
I investigate the case of
non-negligible covariance $\rho$ between the signals.  
Noise in the correlation can increase
or decrease, depending on quantizer parameters,
when $\rho$ increases.
This contrasts with correlation of
continuously-valued or unquantized signals, for which the 
noise in phase with $\rho$ 
increases with increasing $\rho$,
and noise out of phase decreases.  Indeed, for some quantizer parameters, 
I find that correlation of quantized signals provides a more accurate
estimate of $\rho$ than would correlation without
quantization.  
I present analytic results in exact form and as 
polynomial expansions, and compare these mathematical results
with results of computer simulations.

\end{abstract}


\section{INTRODUCTION}

Nearly all signals from astrophysical sources can be represented
as electric fields comprised of Gaussian noise.  These noiselike
signals have zero
mean.  All information about the source is contained in the variance
of the electric field, and in covariances between different polarizations,
positions, times, or frequency ranges.  
The intensity, for example, is simply the sum of
the variances in 2 basis polarizations.  More
generally, all the Stokes parameters can be expressed in terms of
the variances and covariances of of these 2 polarizations.  Similarly,
in interferometry, the covariance of electric fields 
at different positions is the visibility, the Fourier transform of 
source structure.  In correlation spectroscopy, the
covariances of electric field at different time separations, expressed as the
autocorrelation function, are the Fourier transform of 
the spectrum.  Because the signals are drawn from
Gaussian distributions, 
their variances and covariances completely characterize them.  
The single known exception to this rule of Gaussian statistics
is radiation from pulsars, 
under certain observing conditions \citep{jen01}.

Particularly at wavelengths of a millimeter or more,
covariances are usually estimated by correlation.
(Actually correlation is used at all wavelengths,
but at wavelengths shortward of a millimeter quantum-mechanical
processes come into the picture, complicating it).
Correlation involves forming products of samples of the two signals.
The average of many such products approximates the
covariance.
In mathematical terms,
for two signals $x$ and $y$,
the covariance is $\rho = \langle x y\rangle$,
where the angular brackets $\langle ..\rangle$ represent
a statistical average over an ensemble of all
statistically-identical signals.
Correlation approximates this enormously infinite average
with a finite average over $N_q$ samples of $x$ and $y$:
$r_{\infty} = {{1}\over{N_q}}\sum_{i=1}^{N_q} x_i y_i$.
Here, the subscript ``$\infty$'' reflects the fact that $x$ and $y$
are unquantized; their accuracy is not limited to a finite
number of quantized levels.
The subscript ``$q$'' indicates sampling at the Nyquist
rate, as I assume (see \citet{tms86}).

Because the number of samples in most measurements of correlation
is large,
the results of a finite correlation follow a Gaussian distribution.
This is a consequence of the central limit theorem.
Thus, one expects a set of identical measurements of $r_{\infty}$ to 
be fully characterized by their mean, $\langle r_{\infty}\rangle$,
and their standard deviation, 
$\sqrt{\langle (r_{\infty}-\langle r_{\infty}\rangle)^2\rangle}$.
The mean is the deterministic part of the measurement;
it provides an estimate of $\rho$.
The standard deviation characterizes the random part of the measurement,
and is often called ``noise'' (but is to be distinguished from
the noiselike signals $x$ and $y$ that are being correlated).
In principle,
the best measurement minimizes the random part,
while preserving the relation between the deterministic
part and $\rho$.
The signal-to-nose ratio of the correlation,
${\cal R_{\infty}}=\langle r_{\infty}\rangle/\sqrt{\langle (r_{\infty}-\langle r_{\infty}\rangle)^2\rangle}$,
provides a figure of merit that 
quantifies the relative sizes of deterministic and
random parts; see, for example, \citet{tms86}.

The electric field is commonly digitized before correlation.
Digitization includes sampling and quantization.  Sampling involves
averaging the signal over short time windows; it thus restricts the
range of frequencies that can be uniquely represented.
For simplicity, in this paper I will restrict discussion
to ``video'' or ``baseband'' signals,
for which a frequency range of 0 up to some maximum frequency is present;
and that they are sampled at the Nyquist rate, or at half the shortest period
represented.  
I also assume that the signals are ``white,'' in the sense
that samples $x_i$ and $y_j$ are correlated only if $i=j$;
and that the signals are stationary, so that
the correlation of $x_i$ and $y_j$ is independent of $i$.
These assumptions limit the influence of sampling.
I will discuss spectrally-varying signals elsewhere \citep{gwi03}.

Quantization limits the values that can be represented,
so that the digitized signal imperfectly represents the 
actual signal.
Quantization thus introduces changes in both the mean
correlation $\langle r_M\rangle$ and its standard deviation 
$\sqrt{\langle (r_M-\langle r_M\rangle)^2\rangle}$.
Here the subscript ``$M$'' represents the fact that the 
quantized signal can take on $M$ discrete values.
The mean and standard deviation of $r_M$
can be calculated from the statistics of the quantized
signals $\hat x_{i}$ and $\hat y_{i}$ and the details
of the quantization scheme.

A number of previous authors have addressed the effects of 
quantization on correlation of noiselike signals
(see, for example, \citet{tms86}, Chapter 8, and references therein).
Notably, \citet{van66} found the average
correlation and the standard deviation
for two-level quantization:
the case where quantization reduces the signals to only their signs.
\citet{coo70} found the average correlation
and its standard deviation,
for small normalized covariance $\rho<<1$,
for four-level correlation:
quantization reduces the signals to signs
and whether they lie above or below some threshold $v_0$.
He found the optimal values of $v_0$,
and the relative weighting of points above and below the threshold $n$,
as quantified by the signal-to-noise ratio ${\cal R}_4$.
\citet{hag73} generalized this to a broader range of
quantization schemes, and studied effects of oversampling.
\citet{bow74} examined Gaussian noise and a signal of general form
in the small-signal limit.
They devise a criterion for the accuracy with which a quantization
scheme represents a signal, and
show that this yields the highest signal-to-noise ratio for $\rho<<1$.
Most recently,
\citet{jen98} examined the case of many-level correlators.
They use 
a criterion similar to that of \citet{bow74} to
calculate the optimal level locations for various numbers of levels.
\citet{jen98} also find the mean spectrum for a spectrally-varying
source, as measured by an autocorrelation spectrometer;
they find that quantization introduces a uniform offset to to the spectrum,
and scales the spectrum by a factor, and calculate the offset and factor.
\citet{dad84} provide an extensive analysis of errors in a 3-level correlator.

In this paper, I calculate the average correlation and
its standard deviation for nonvanishing covariance $\rho$.
I provide exact expressions for these quantities,
and approximations valid
through 
fourth order in $\rho$.
Interestingly,
noise actually declines for large $\rho$, 
for correlation of quantized signals.
Indeed, the signal-to-noise ratio
for larger $\rho$ can actually exceed that for 
correlation of an unquantized signal.
In other words,
correlation of a quantized signal
can provide a more accurate measure of
$\rho$ than would correlation of the signal before quantization.
This fact is perhaps surprising;
it reflects the fact that correlation 
is not always
the most accurate way to determine
the covariance of two signals.

The organization of this paper is as follows:
In section 2, I review the statistics of
the correlation of unquantized, or continuously variable, complex signals.
In section 3, I present expressions that give the
average correlation and the standard deviation,
in terms of integrals involving the 
characteristic curve.
I include statistics of real and imaginary parts,
which are different.
I present expansions of these integrals 
as a power series in $\rho$.
In section 4 I discuss
computer simulations of correlation
to illustrate these mathematical results.
I summarize the results in the final section.

\section{CORRELATION OF UNQUANTIZED VARIABLES}\label{continuous_correlation}

\subsection{Bivariate Gaussian Distribution}

Consider two random, complex signals $x$ and $y$.
Suppose that each of these signals is 
a random variable drawn from a Gaussian distribution.
Suppose that the signals $x$ and $y$ are correlated,
so that they are, in fact, drawn from a joint Gaussian
joint probability density function $P(x,y)$ \citep{mey75}.
Without loss of generality, I assume that each signal
has variance of 2:
\begin{equation}
\langle x x^*\rangle = \langle y y^*\rangle \equiv 2.
\end{equation}
In this expression, the angular brackets
$\langle ... \rangle$ denote an statistical average:
in other words, an average over all systems with the specified statistics.
This choice for a variance of 2 for $x$ and $y$ is
consistent with the
literature on this subject,
much of which treats real signals (rather than complex ones),
drawn from Gaussian distributions with unit variance
\citep{coo70, tms86}.
Of course, the results presented here
are easily scaled to other variances for the input signal.
I demand that the signals themselves have no intrinsic
phase: in other words, the statistics
remain invariant under the transformation
$x\rightarrow x e^{i\phi_0}$, and
$y\rightarrow y e^{i\phi_0}$,
where $\phi_0$ is an arbitrary overall phase.
It then follows that
\begin{equation}
\langle x x\rangle = \langle y y\rangle=0.
\end{equation}
From these facts, one finds:
\begin{eqnarray}
\langle \re [x] \re [x] \rangle &=&\langle \im [x] \im [x] \rangle =1 \label{re_im_sigmas}\\
\langle \re [x] \im [x] \rangle &=& 0, \nonumber
\end{eqnarray}
and similarly for $y$.
Thus, real and imaginary parts are drawn from Gaussian distributions
with unit variance.  The distributions are circular in the complex plane for
both $x$ and $y$.

Without loss of generality,
I assume that the normalized covariance of the signals $\rho$ is purely
real:
\begin{equation}
\rho \equiv {{\langle x y^*\rangle}\over{\sqrt{\langle |x|^2\rangle \langle |y|^2\rangle}}}
=
\haf \langle x y^*\rangle .
\end{equation}
One can always make $\rho$ purely real by 
rotating $x$ (or $y$) in the complex plane:
$x\rightarrow x e^{i\phi_x}$, and
$y\rightarrow y$.
Note that,
because of the absence of any intrinsic phase,
\begin{eqnarray}
\langle \re [x] \re [y] \rangle &=&\langle \im [x] \im [y] \rangle =\rho \label{re_im_rho}\\
\langle \re [x] \im [y] \rangle &=&\langle \im [x] \re [y] \rangle =0, \nonumber
\end{eqnarray}
so that
\begin{equation}
\langle x y\rangle=0.
\end{equation}
In other words,
the real parts of $x$ and $y$ are correlated;
and the imaginary parts of $x$ and $y$ are correlated;
but real and imaginary parts are uncorrelated.
In mathematical terms,
real parts (or imaginary parts) are drawn from the bivariate Gaussian distribution:
\begin{equation}
P_2(X, Y)=
{{1}\over{2\pi\sqrt{1-\rho^2}}}
\exp\left\{ {{-1}\over{2(1-\rho^2)}}( X^2+ Y^2- 2 \rho\, X Y)\right\}, 
\label{bivariate_Gaussian_dist}
\end{equation}
where $(X,Y)$ stands for either $(\re[x], \re[y])$ or $(\im[x],\im[y])$.
The distributions for real and imaginary parts are identical,
but real and imaginary parts are uncorrelated.
Therefore,
\begin{eqnarray}
P\left(x,y\right)&=&
P_2\left(\re [x], \re [y]\right)\times P_2\left(\im [x], \im [y]\right).
\label{product_2_bivariates}
\end{eqnarray}

\subsection{Correlation of Unquantized Signals}

\subsubsection{Correlation}

Consider the product of two random complex signals,
drawn from Gaussian distributions as in the previous section,
sampled in time.
Suppose that the signals are not quantized:
they can take on any complex value.
The product of a pair
of samples $c_i=x_i y_i^*$
does not follow a Gaussian distribution.
Rather, 
the distribution of $c_i$ is the product of an exponential 
of the real part, multiplied by the modified Bessel function
of the second kind of order zero of the magnitude of $c_i$ \citep{gwi01}.
However, the average of many such products, 
averaged over a large number of pairs of samples,
approaches a Gaussian distribution,
as the central limit theorem implies.
In such a large but finite sum,
\begin{equation}
r_{\infty}={{1}\over{2 N_q}} \sum_{i=1}^{N_q} x_{i} y_{i}^*
\label{r_def}
\end{equation}
provides an estimate of the covariance $\rho$.
Here the index $i$ runs over the samples,
commonly samples taken at different times.
The total number of samples correlated is $N_q$.
Henceforth I assume that in all summations,
indices run from $1$ to $N_q$.
The subscript $\infty$ on the correlation $r_{\infty}$
again indicates that the 
correlation has been formed for variables $x_i$, $y_i$ that can
take on any of an infinite number of values;
in other words, it indicates that $x_i$ and $y_i$ 
have not been quantized.

\subsubsection{Mean Correlation for Unquantized Signals}

The mean correlation is equal to the covariance,
in a statistical average:
\begin{equation}
\langle r_{\infty}\rangle =\langle \re[r_{\infty}]\rangle 
={{1}\over{2}}\left(\langle\re[x_i]\re[y_i]\rangle + \langle\im[x_i]\im[y_i]\rangle\right)
=\rho,
\label{r_bar}
\end{equation}
where I used the assumption that
the phase of $\rho$ is zero, Eq.\ \ref{re_im_rho}.
This can also be seen from the mean of the 
distribution of the products $c_i=x_i y_i^*$ \citep{gwi01},
or simply by integrating over the joint distribution
of $x_i$ and $y_i$, Eqs.\ \ref{bivariate_Gaussian_dist}
and\ \ref{product_2_bivariates}.

\subsubsection{Noise for Correlation of Unquantized Signals}

Because the distribution of $r_{\infty}$ is
Gaussian, 
the distribution of $r_{\infty}$
is completely characterized by its mean, Eq.\ \ref{r_bar},
and by the variances $\langle r_{\infty} r_{\infty}^*\rangle$ and
$\langle r_{\infty} r_{\infty}\rangle$.
Suppose that the samples $x_i$ and $y_i$ are
independent;
in mathematical terms,
suppose that 
\begin{equation}
x_i y_j = 0,\quad{\rm for\ }i\neq j .
\end{equation}
If they are not independent the results will depend
on the correlations among samples; 
this case is important if, for example,
the signal has significant spectral structure.
\citet{jen98} find the
average spectrum in this case; I will discuss the noise
in future work.
Here I consider only independent samples.
In that case,
\begin{eqnarray}
\langle r_{\infty} r_{\infty}^*\rangle 
&=&{{1}\over{4 N_q^2}}\sum_{ij}
\langle x_{i} y_{i}^* x_{j}^* y_{j}\rangle \\
&=&{{1}\over{4 N_q^2}}\sum_{i}
\langle x_{i} y_{i}^* x_{i}^* y_{i}\rangle 
+{{1}\over{4 N_q^2}}\sum_{i\neq j}
\langle x_{i} y_{i}^*\rangle \langle x_{j} y_{j}^*\rangle \nonumber \\
&=&{{1}\over{4 N_q}}\langle x_{i} y_{i}^* x_{i}^* y_{i}\rangle
+{{N_q-1}\over{4 N_q}}(2\rho)^2 , \nonumber
\end{eqnarray}
where I have
separated the terms with $i=j$
from those with $i\neq j$,
and appealed to the facts that
$x_i$ and $y_j$ are covariant only if $i=j$ (``white'' signals),
and that for $i=j$ the statistics are stationary in $i$.
For Gaussian variables with zero mean
$a,\, b,\, c,\, d$,
all moments are related to the second moments:
\begin{equation}
\langle a b c d\rangle 
=\langle a b\rangle \langle c d\rangle 
+\langle a c\rangle \langle b d\rangle 
+\langle a d\rangle \langle b c\rangle ,
\label{Gaussian_4th_moment_exp}
\end{equation}
so that:
\begin{eqnarray}
\langle x_{i} y_{i}^* x_{i}^* y_{i}\rangle &=& (2\rho)^2 + 4 .
\end{eqnarray}
Therefore,
\begin{eqnarray}
\langle r_{\infty} r_{\infty}^* \rangle 
&=&{{1}\over{N_q}} +\rho^2 \label{rr*} \\
\langle r_{\infty} r_{\infty}^* \rangle - \langle r_{\infty} \rangle \langle r_{\infty}^* \rangle 
&=& {{1}\over{N_q}} \nonumber
\end{eqnarray}
An analogous calculation yields
\begin{equation}
\langle x_{i} y_{i}^* x_{i} y_{i}^*\rangle = 2 (2\rho)^2 ,
\end{equation}
and so
\begin{eqnarray}
\langle r_{\infty}^2 \rangle &=& {{1}\over{N_q}} \rho^2 +\rho^2 \label{rr} \\
\langle r_{\infty}^2 \rangle - \langle r_{\infty}\rangle^2 &=& {{1}\over{N_q}} \rho^2 \nonumber
\end{eqnarray}
I combine these facts to find
the means and standard deviations of the real and imaginary parts of
the measured correlation, $r_{\infty}$:
{ \begin{eqnarray}
\langle\re [r_{\infty}]\rangle&=&\langle r_{\infty}\rangle
\phantom{\haf\left\{\langle r_{\infty}r_{\infty}^*\rangle+\langle r_{\infty}r_{\infty}\rangle\right\}-^2}=
\rho 
\label{r_infty_avgsd} \\
\langle\im [r_{\infty}]\rangle&& 
\phantom{\haf\left\{\langle r_{\infty}r_{\infty}^*\rangle+\langle r_{\infty}r_{\infty}\rangle\right\}-\langle r_{\infty}\rangle^2}=
0  \nonumber \\
\langle\re[r_{\infty}]^2\rangle - \langle \re [r_{\infty}]\rangle^2 &=& 
\haf\left\{\langle r_{\infty}r_{\infty}^*\rangle+\langle r_{\infty}r_{\infty}\rangle\right\}-\langle r_{\infty}\rangle^2=
{{1}\over{2 N_q}}(1+\rho^2) \nonumber \\
\langle\im [r_{\infty}]^2\rangle
&=&\haf\left\{\langle r_{\infty}r_{\infty}^*\rangle-\langle r_{\infty}r_{\infty}\rangle\right\}
\phantom{ - \langle r_{\infty}\rangle^2 }
={{1}\over{2 N_q}}(1-\rho^2)\nonumber .
\end{eqnarray} }

If the number of independent samples $N_q$ is large,
the central limit theorem implies that 
$\re [r_{\infty}]$ and $\im [r_{\infty}]$ are drawn from Gaussian distributions.
The means and variances of these distributions,
as given by Eq.\ \ref{r_infty_avgsd},
completely characterize $r_{\infty}$.
The fact that the real part of $r_{\infty}$ 
has greater standard deviation than the imaginary part
reflects the presence of self-noise or source-noise.
Sometimes this is described as the contribution of the noiselike
signal to the noise in the result.

Commonly, and often realistically,
astrophysicists suppose that $\rho$ measures the intensity of 
one signal
that has been superposed with two uncorrelated noise signals 
to produce $x$ and $y$
(see \citet{bow74,kul89,ana91}).
A change in $\rho$
then corresponds to a change in the intensities
$\langle |x|^2\rangle$ and $\langle |y|^2\rangle$ as well.
Here, I suppose that $\langle |x|^2\rangle =\langle |y|^2\rangle \equiv 1$,
while $\rho$ varies.
The results presented here can be scaled to those for the alternative interpretation.

\subsubsection{SNR for Correlation of Unquantized Signals}

The signal-to-noise ratio (SNR) for $\re [r_{\infty}]$ is:
\begin{equation}
{\cal R}_{\infty}(\re [r_{\infty}])
={{\langle \re [r_{\infty}]\rangle}\over
{\sqrt{\langle \re [r_{\infty}]^2\rangle-\langle \re [r_{\infty}]\rangle^2}}}
={\sqrt{2 N_q}}{{\rho}\over{\sqrt{1 + \rho^2}}}.
\label{SNRinfty}
\end{equation}
Note that for a given number of observations $N_q$,
SNR increases with $\rho$;
the increase is proportional for $\rho<<1$.

A related quantity to SNR is the rms phase,
statistically averaged over many measurements.
The phase is $\phi[r_{\infty}]=\tan^{-1}(\im [r_{\infty}]/\re [r_{\infty}])$.
When the number of observations is sufficiently large,
and the true value of the phase is $0$, as assumed here,
the standard deviation of the phase is 
$\langle \phi[r_{\infty}]^2\rangle 
= \sqrt{ \langle \im [r_{\infty}]^2 \rangle }/\langle \re [r_{\infty}]\rangle $.
The inverse of the standard deviation
of the phase (in radians)
is analogous to the SNR for the real part, Eq.\ \ref{SNRinfty}.
This SNR for phase is:
\begin{equation}
{\cal R}_{\infty}(\phi [r_{\infty}]) = 
{{\langle \re [r_{\infty}]\rangle }\over{\sqrt{\langle \im [r_{\infty}]^2\rangle}}}
={\sqrt{2 N_q}} {{\rho}\over{\sqrt{1-\rho^2}}} .
\label{SNRphase}
\end{equation}
For constant $N_q$, the SNR of the phase increases with $\rho$,
and increases much faster than proportionately for 
$\rho\rightarrow 1$.

%
%

\section{QUANTIZED SIGNALS}\label{quantized_signals}

\begin{figure}[t]
\plotone{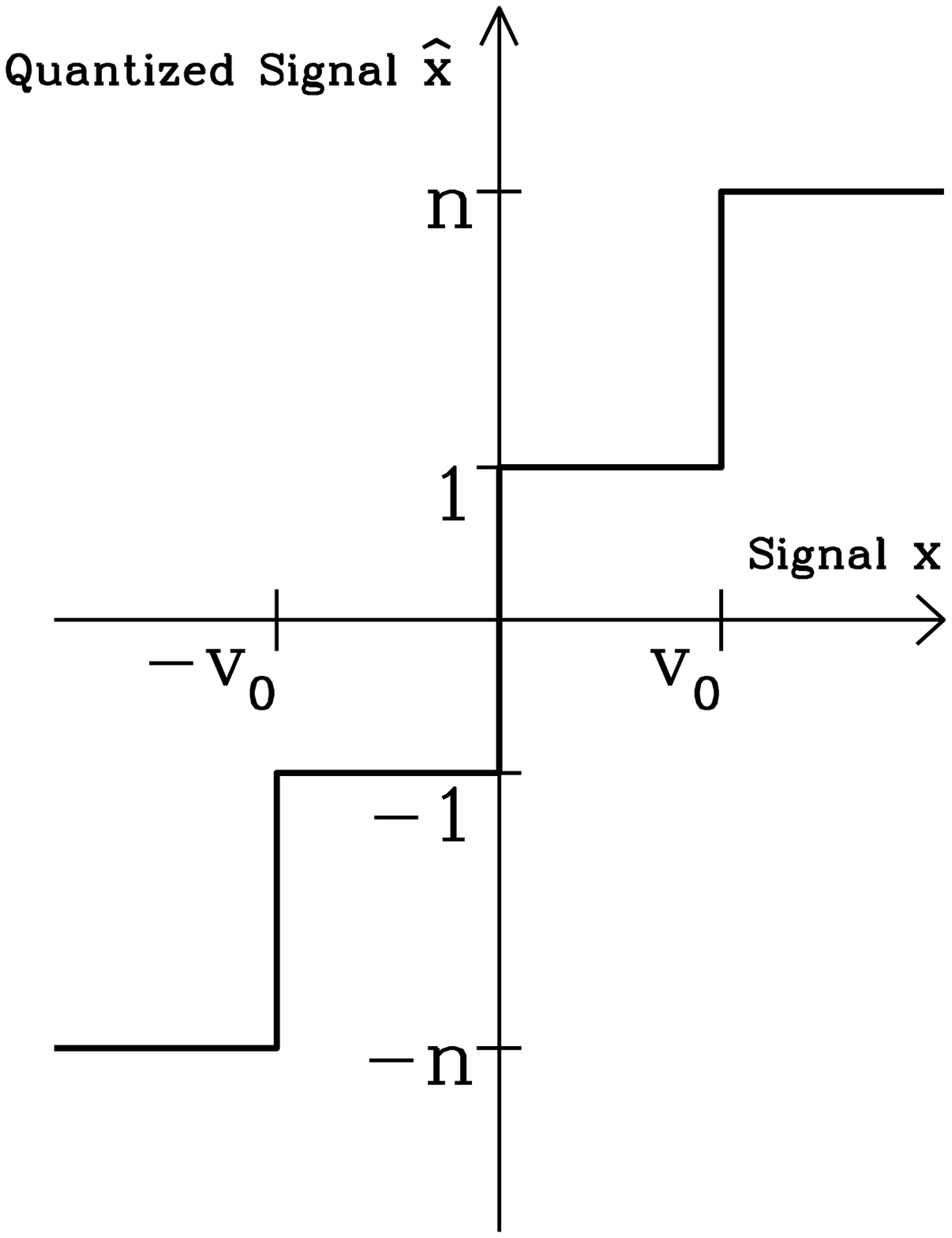}
\figcaption[]{
Characteristic curve $\hat X(X)$ for 4-level quantization.
\label{4_level}}
\end{figure}

Quantization converts 
the variables $x$, $y$
to discrete variables $\hat x$, $\hat y$.
These discrete variables depend on $x$ and $y$
through a multiple step function,
known as a characteristic curve.
Each step extends over some range $[v_{i},v_{i+1}]$ of the unquantized signal,
and is given some weight $n_i$ in correlation.
The function $\hat X(X)$ denotes the characteristic curve,
where again $X$ stands for either $\re [x]$ or $\im [x]$.
The complex quantized variable $\hat x$ 
is thus given by $\hat x = \hat X(\re [x]) + i \hat X(\im[x])$.
The same characteristic curve is
applied to real and imaginary parts.
I hold open the possibility that 
the characteristic curves $\hat X(X)$ and $\hat Y(Y)$ are different.
I assume in this paper that the characteristic curves
are antisymmetric: $\hat X(-X)=-\hat X(X)$, and similarly for $Y$.
\citet{dad84} describe effects of departures from antisymmetry,
and how antisymmetry can be enforced.
Figure \ref{4_level} 
shows a typical characteristic curve for 4-level (or 2-bit) sampling.

Systems with $M$ levels of quantization
can be described by analogous, more complicated characteristic curves,
and corresponding sets of weights $\{n_{i}\}$ and
levels $\{v_{i x}\}$ and $\{v_{i y}\}$ \citep{jen98}.
For $M$-level sampling, the correlation of $N_q$ quantized samples is 
\begin{equation}
\hat r_M={{1}\over{2 N_q}} 
\sum_{\imath=1}^{N_q}\hat x_{\imath} \hat y_{\imath}^* .
\end{equation}

In practical correlators,
deviations from theoretical performance
can often be expressed as deviations of the characteristic curve
from its desired form.
In principle, these can be measured by counting the numbers of samples in
the various quantization ranges, and using the assumed Gaussian distributions of the input
signals to determine the actual levels $v_{ix}$ and $v_{iy}$.
\citet{dad84} present an extensive discussion of such errors, 
and techniques to control and correct them.

Although the results presented below are applicable
to more complicated systems,
the 4-level correlator will be used as a specific example in this paper,
with correlation $\hat r_4$.
For 4-level sampling,
commonly 
a sign bit gives the sign of $X$, and an amplitude bit assigns
weight $1$ if $|X|$ is less than some threshold $v_{0x}$,
and weight $n$ if $|X|$ is greater than $v_{0x}$.
Together, sign and amplitude bits describe the 
4 values possible for $\hat X(X)$.
Other types of correlators,
including 2-level, 3-level, or ``reduced'' 4-level (in which
case 
the smallest product, for $|\hat X|=1$, $|\hat Y|=1$,
is ignored),
can be formed
as special cases or sums of four-level correlators \citep{hag73}.

\subsection{Correlation of Quantized Signals: Exact Results}\label{exact_statistics}

\subsubsection{Mean Correlation: Exact Result}

Ideally, from measurement of of the quantized correlation $\hat r_M$ 
one can estimate the true covariance $\rho$.
The statistical mean of $\hat r_M$ is:
\begin{eqnarray}
\langle \hat r_M\rangle &=& {{1}\over{2 N_q}}\sum_i \langle \hat x_i \hat y_i^*\rangle 
\label{r_M_avg} \\
&=&
\haf \langle 
( \re [\hat x] \re [\hat y] +\im [\hat x]\im [\hat y] )+
i
(-\re [\hat x] \im [\hat y] +\im [\hat x]\re [\hat y] ) 
\rangle .  \nonumber
\end{eqnarray}
Because $\re [\hat x]$ depends only on $\re[x]$ and $\im [\hat y]$ depends only on $\im[y]$,
and $\re[x]$ and $\im[y]$ are completely independent
(and similarly for $\im[\hat x]$ and $\re[\hat y]$),
\begin{equation}
\langle \re[\hat x]\im [\hat y]\rangle=\langle \im[\hat x]\re [\hat y]\rangle=0.
\end{equation}
Thus, the imaginary part of $\langle\hat r\rangle$,
which involves products of these statistically independent terms, has average zero (Eq.\ \ref{re_im_sigmas}).
For the real part, 
\begin{equation}
\langle \re [\hat x] \re [\hat y]\rangle = \langle \im [\hat x]\im [\hat y]\rangle ,
\end{equation}
where I use the assumption that the characteristic curves are identical 
for real and imaginary parts,
and that real and imaginary parts of $x$ and $y$ have identical statistics
(Eqs.\ \ref{re_im_sigmas}, \ref{re_im_rho}).
I use the bivariate Gaussian distribution
for real and imaginary parts to find a formal expression
for the statistical average of the correlation:
\begin{equation}
\langle \hat r_M\rangle = \langle \re [\hat x] \re [\hat y]\rangle = \Upsilon_{XY}\equiv \int dX dY P_2(X,Y) \hat X(X)\hat Y (Y). \label{r_M_exact}
\end{equation}
This integral defines $\Upsilon_{XY}$.
For the assumed antisymmetric characteristic curves $\hat X(X)$ and $\hat Y(Y)$,
one can easily show that
${{\partial \Upsilon_{XY}}\over {\partial \rho}}>0$.
In other words, the ensemble-averaged quantized correlation is an increasing function
of the covariance $\rho$,
for completely arbitrary quantizer settings 
(so long as the characteristic curves are antisymmetric).

The discussion of this section reduces the 
calculation of the average quantized correlation
to that of integrating $P_2(X,Y)$ 
over each rectangle in a grid, with the edges of the rectangles given by the thresholds in the
characteristic curves \citep{kok01}.
The function $\Upsilon_{XY}$ 
depends on $\rho$ through $P_2(X,Y)$.
This function is usually expanded through first order in $\rho$,
because $\rho$ is small in most astrophysical observations.

\subsubsection{Simpler Form for $\Upsilon_{XY}$}\label{simplerform}

The integral $\Upsilon_{XY}$ and similar integrals can be
converted into one-dimensional integrals for easier analysis.
If one defines
\begin{equation}
Q(v_{0x},v_{0Y})=
\int_{v_{0X}}^\infty dX \int_{v_{0Y}}^\infty dY
P_2(X,Y) ,
\end{equation}
then the Fourier transform of 
$P_2(v_{0x},v_{0Y})$ is equal to that of $\partial Q(v_{0x},v_{0Y})/\partial \rho$,
as one finds from integration by parts.
Thus,
\begin{equation}
Q(v_{0x},v_{0Y}) = \int_0^{\rho} d\rho_1 P_2(v_{0x},v_{0Y}) 
+ \qua {\rm erfc}({{v_{0X}}\over{\sqrt{2}}})
\,{\rm erfc}({{v_{0Y}}\over{\sqrt{2}}}). \nonumber
\end{equation}
The integral $\Upsilon_{XY}$ is the sum of one such integral and
one such constant for each step in the characteristic curve.
This one-dimensional form is useful for numerical evaluation and expansions.
\citet{Kas01} also present an interesting expansion of $\Upsilon_{XY}$ 
in Hermite polynomials, in $v_{0}$.

\subsubsection{Noise: Exact Results}

This section presents
an exact expression for the variance
of the correlation of a quantized signal,
when averaged over the ensemble of all statistically identical
measurements.
Real and imaginary parts of
$\hat r_M$
have different variances.
This requires calculation of both $\langle \hat r_M \hat r_M^* \rangle$
and $\langle \hat r_M \hat r_M \rangle$.  
Note that:
\begin{eqnarray}
\langle \hat r_M \hat r_M^* \rangle &=&{{1}\over{4 N_q^2}}\langle \sum_{i} \sum_{j} \hat x_i \hat y_i^* \hat x_j^* \hat y_j\rangle \label{r_M_r_M*}\\
&=&{{1}\over{4 N_q^2}}\sum_{i}\langle \hat x_i \hat x_i^* \hat y_i \hat y_i^* \rangle 
 + {{1}\over{4 N_q^2}} \sum_{i\neq j}\langle \hat x_i \hat y_i^*\rangle\langle \hat x_j^* \hat y_j \rangle \nonumber \\
&=& {{1}\over{4 N_q}} \langle \hat x_i \hat x_i^* \hat y_i \hat y_i^* \rangle +
{{N_q-1}\over{N_q}} \langle \hat r_M \rangle^2 \nonumber
\end{eqnarray}
where the sum over $i\neq j$ is simplified by the fact that samples at different times are uncorrelated,
and by Eq.\ \ref{r_M_avg}.
I expand the first average in the last line:
\begin{eqnarray}
\langle \hat x_i \hat x_i^* \hat y_i \hat y_i^* \rangle &=&
\langle \re [\hat x_i]^2 \re [\hat y_i]^2 + \im [\hat x_i]^2 \im [\hat y_i]^2 + \re [\hat x_i]^2 \im [\hat y_i]^2 + \im [\hat x_i]^2 \re [\hat y_i]^2 \rangle \label{xi_xis_yi_yis} \\
&=&\langle \re [\hat x_i]^2 \re [\hat y_i]^2\rangle + \langle\im [\hat x_i]^2 \im [\hat y_i]^2 \rangle
+ \langle \re [\hat x_i]^2\rangle\langle \im [\hat y_i]^2\rangle + \langle\im [\hat x_i]^2 \rangle\langle\re [\hat y_i]^2\rangle , \nonumber
\end{eqnarray}
where I have used the fact that the real part of $\hat x$ has zero covariance with the
imaginary part of $\hat y$, and vice versa.
Because the real and imaginary parts are identical,
this sum can be expressed formally in terms of the integrals:
\begin{eqnarray}
\langle \re [\hat x_i]^2 \re [\hat y_i]^2\rangle =  \langle\im [\hat x_i]^2 \im [\hat y_i]^2 \rangle &=& 
\Upsilon_{X2Y2}\equiv 
\int dX\, dY \; P(X,Y) \; \hat X(X)^2 \; \hat Y(Y)^2 , \label{UpsX2Y2_AX2_AY2} \\
\langle\re[\hat x_i]^2\rangle=\langle\im[\hat x_i]^2\rangle &=& A_{X2} \equiv 
\int dX {{1}\over{\sqrt{2\pi}}} e^{-\haf X^2} \hat X(X)^2   \nonumber \\
\langle \re [\hat y_i]^2\rangle =\langle\im [\hat y_i]^2\rangle &=& A_{Y2} \equiv 
\int dY {{1}\over{\sqrt{2\pi}}}e^{-\haf Y^2} \hat Y(Y)^2 . \nonumber
\end{eqnarray}
These expressions
defines $\Upsilon_{X2Y2}$, $A_{X2}$, and $A_{Y2}$.
Thus,
\begin{eqnarray}
\langle \hat r_M \hat r_M^* \rangle &=&{{1}\over{N_q}}\left\{\haf\Upsilon_{X2Y2}+\haf A_{X2}A_{Y2}-(\Upsilon_{XY})^2 \right\}
+ (\Upsilon_{XY})^2 . \label{r_M_r_M*_exact}
\end{eqnarray}
Note that in these expressions 
$A_{X2}$ and $A_{Y2}$ are constants that depend on the characteristic curve, but not on $\rho$;
whereas $\Upsilon_{XY}$ and
$\Upsilon_{X2Y2}$ depend on $\rho$ in complicated ways, 
as well as on the characteristic curve.

Similarly,
\begin{eqnarray}
\langle \hat r_M \hat r_M \rangle &=& {{1}\over{4 N_q}} \langle \hat x_i \hat y_i^* \hat x_i \hat y_i^* \rangle +
{{N_q-1}\over{N_q}} \langle \hat r_M \rangle^2 .
\end{eqnarray}
I again expand the first sum in the last line:
\begin{eqnarray}
\langle\hat x_i\hat y_i^*\hat x_i\hat y_i^*\rangle 
&=& 
2 \Upsilon_{X2Y2} - 2 A_{X2}A_{Y2} + 4 \Upsilon_{XY}^2 . 
\label{expand_4part_sum}
\end{eqnarray}
where I omit the imaginary terms, all of which average to zero.
Therefore:
\begin{eqnarray}
\langle \hat r_M \hat r_M \rangle &=& {{1}\over{N_q}} \left( \haf \Upsilon_{X2Y2} - \haf A_{X2}A_{Y2} \right) 
+ (\Upsilon_{XY})^2 . \label{r_M_r_M_exact}
\end{eqnarray}

\begin{figure}[t]
\plotone{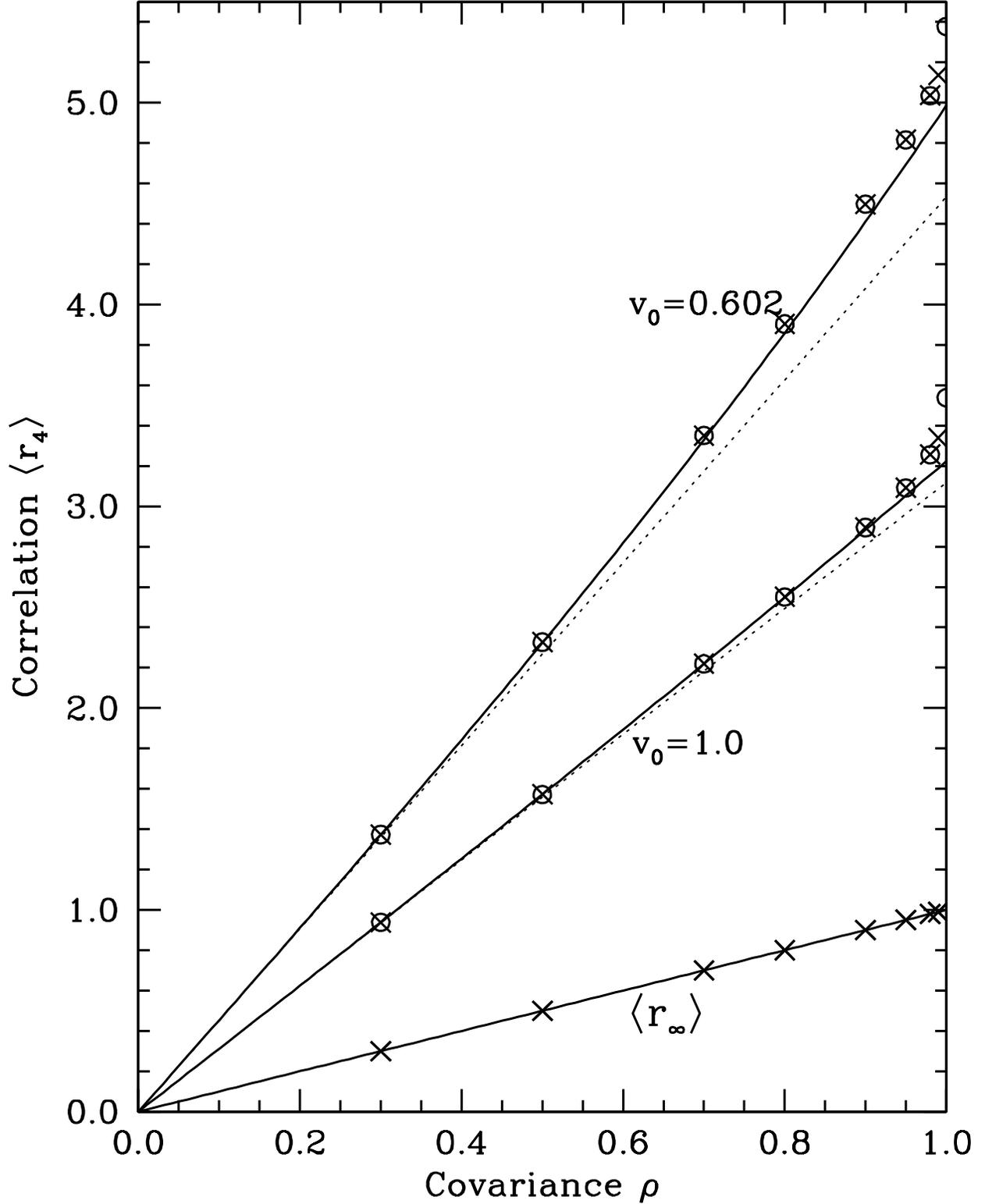}
\figcaption[]{
Average correlation plotted with covariance $\rho$.
Curves show $\langle r_{\infty}\rangle$,
and approximations to
$\langle\hat r_4\rangle$ 
for $v_0=1.0$ and $v_0=0.602$,
with $n=3$.
Heavy lines show the third-order approximation of Eq.\ \ref{hat_rM_bar_approx},
and light lines show the linear approximation of \citet{coo70}.
Circles 
show true values as computed by direct integration of Eq.\ \ref{r_M_exact}.
Crosses show results of simulations (\S \ref{simulations}).
The sharp
rise in  $\langle\hat r_4\rangle$ 
for $\rho\approx 1$, shown by the circles at far right,
motivates the approximation of \citet{jen98} that $\langle\hat r_4\rangle$
varies proportionately with $\rho$ for $\rho<1$,
with a spike at $\rho=1$.
\label{re_plot}}
\end{figure}

\begin{figure}[t]
\plotone{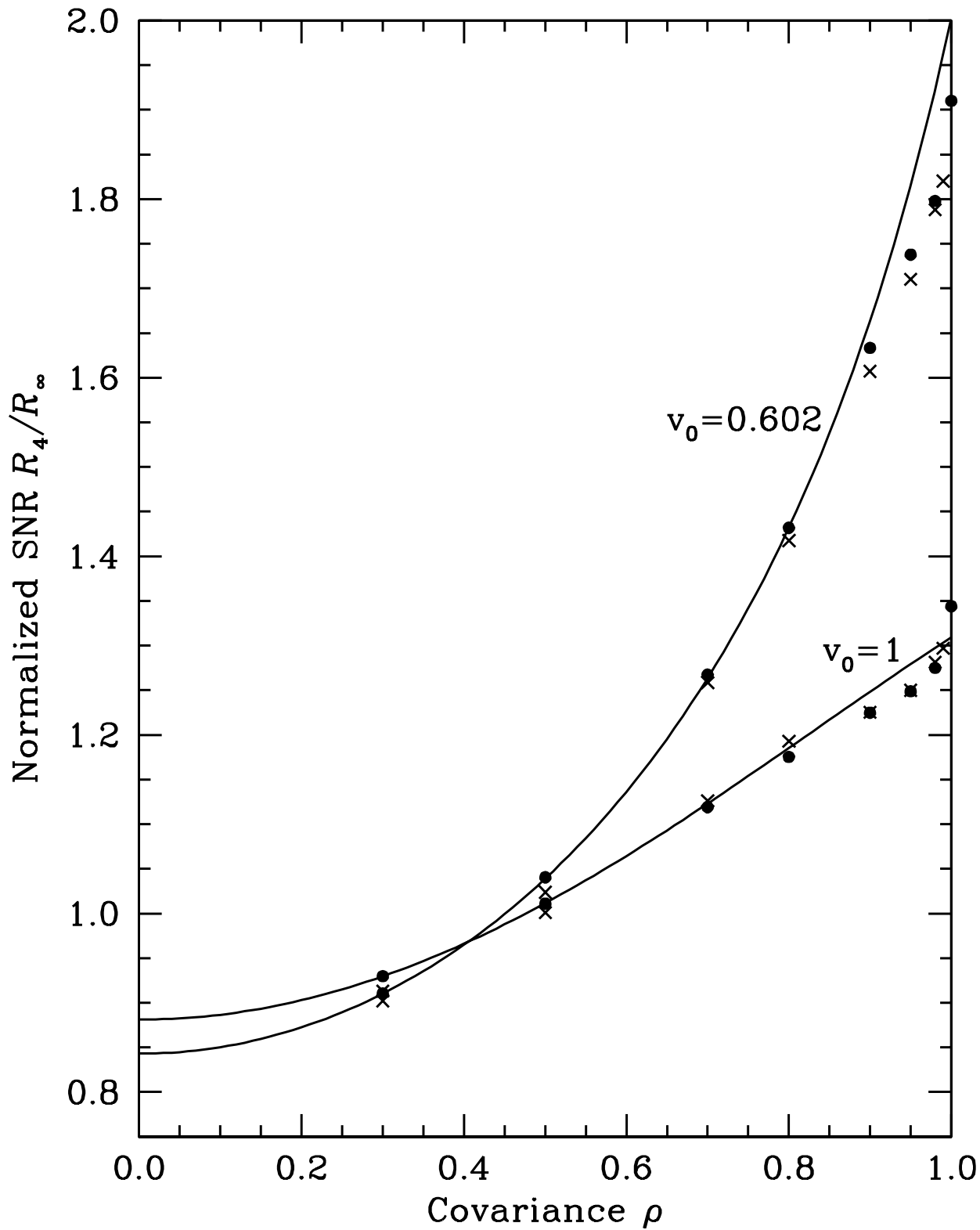}
\figcaption[]{
SNR for quantized signals, normalized to SNR
of the unquantized signal:
${\cal R}_{4}/{\cal R}_{\infty}$,
plotted with covariance $\rho$.
Curves mark the 
approximate expressions given by Eqs.\ \ref{hat_rM_bar_approx} and\ \ref{hat_rM_sd_approx}
for ${\cal R}_4$,
normalized by ${\cal R}_{\infty}$
as given by Eq.\ \ref{r_infty_avgsd}.
Dots give exact expressions as found from direct integration of Eq.\ \ref{r_M_exact}.
The y-intercept for $v_0=1.0$ is the standard SNR for a 4-level correlator,
${\cal R}_4=0.88115$, with $v_0=1.0$ and $n=3$,
which is optimal for $\rho=0$.
Note that the ratio can be greater than 1;
this indicates that quantized correlation can
be more efficient than correlation of the original, unquantized signals.
\label{ratio_plot}}
\end{figure}

Using the same logic as in the derivation of Eq.\ \ref{r_infty_avgsd},
Eqs.\ \ref{r_M_r_M*_exact} and\ \ref{r_M_r_M_exact} can be used to find
the means and standard deviations of the real and imaginary parts of $\hat r_M$:
\begin{eqnarray}
\langle \re [\hat r_M]\rangle &=& \Upsilon_{XY} \label{sds_re_im_r_M} \\
\langle \im [\hat r_M]\rangle &=& 0 \nonumber \\
\langle \re [\hat r_M]^2\rangle - \langle \re [\hat r_M]\rangle^2  
&=& {{1}\over{2 N_q}}\left(\Upsilon_{X2Y2}-(\Upsilon_{XY})^2\right)\nonumber \\
\langle \im [\hat r_M]^2\rangle
&=& {{1}\over{2 N_q}}\left(A_{X2}A_{Y2} -(\Upsilon_{XY})^2\right) . \nonumber
\end{eqnarray}
Again, note that
$A_{X2}$ and $A_{Y2}$ are constants that depend on the characteristic curve, 
but not on the covariance $\rho$,
whereas
$\Upsilon_{XY}$ and $\Upsilon_{X2Y2}$ depend on
the actual value of $\rho$ as well as the characteristic curve.
For particular characteristic curves, and particular values of $\rho$,
these expressions nevertheless yield the mean correlation,
and the standard deviations of real and imaginary parts about the mean.
Figures\ \ref{re_plot} through\ \ref{ratio_plot} show examples,
and compare them with the approximate results from the following section.

Note that, because $\Upsilon_{XY}$ is an increasing function of $\rho$,
the standard deviation of the imaginary part decreases with increasing 
covariance $\rho$.
This holds for arbitrary quantizer parameters,
so long as the characteristic curves are antisymmetric.
In other words, the noise in the imaginary part always decreases 
when the correlation increases.

For uncorrelated signals, $\rho=0$.
One finds then that
$\Upsilon_{XY}= 0$
and $\Upsilon_{X2Y2}= A_{X2}A_{Y2}$.
In this case both real and imaginary parts have identical
variances, as they must:
\begin{equation}
\langle \re [\hat r_M]^2\rangle - \langle \re [\hat r_M]\rangle^2  
= 
\langle \im [\hat r_M]^2\rangle =
{{1}\over{2 N_q}}A_{X2}A_{Y2} ,\quad {\rm for\ }\rho=0.
\end{equation}
This recovers the result of \citet{coo70} and others
for the noise.

If the characteristic curves are identical, so that: $\hat X(X)=\hat Y(Y)$,
then if the signals are identical: $\rho= 1$,
one finds that
$\Upsilon_{X2Y2}= A_{X2}= A_{Y2}$.
Under these assumptions then
$\langle \im [\hat r_M]^2\rangle =0$.

\subsection{Correlation of Quantized Signals: Approximate Results}

\subsubsection{Mean Correlation: Approximate Result}\label{mean_approx_sec}

Unfortunately the expressions for the mean correlation and the noise,
for quantized signals,
both depend in a complicated way on the covariance $\rho$,
the quantity one seeks to measure.
Often the covariance $\rho$ is small.
Various authors discuss
the correlation $\hat r_M$ of quantized signals $\hat x$ and $\hat y$
to first order in $\rho$,
as is appropriate in the limit $\rho\rightarrow 0$.
\citep{van66,coo70,hag73,tms86,jen98}.
\citet{jen98} also calculate $\langle r_4\rangle$ for $\rho=1$;
as they point out,
this case is important for autocorrelation spectroscopy.
\citet{dad84} present an expression
for $\langle \hat r\rangle$ for a 3-level correlator,
and present several useful approximate expressions for 
the inverse relationship $\rho(\langle \hat r_M\rangle )$.
Here I find the mean correlation $\langle \hat r_M\rangle$
through fourth order in $\rho$.

For small covariance $\rho$, one can expand 
$P(X,Y)$ in Eq.\ \ref{bivariate_Gaussian_dist}
as a power series in $\rho$:
\begin{eqnarray}
P(X,Y)&=&{{1}\over{2\pi\sqrt{1-\rho^2}}}
\exp\left\{{{-1}\over{2 (1-\rho^2)}}( X^2+ Y^2- 2 \rho X Y)\right\} \label{PDistExpansion} \\
&\approx &{{1}\over{2 \pi}}
e^{-\haf X^2}
e^{-\haf Y^2}
\Big[ 1 + X Y \rho + \haf(1-X^2)(1-Y^2) \rho^2 \nonumber \\
&&\phantom{{{1}\over{2 \pi}} e^{-\haf X^2}e^{-\haf Y^2}  1  }
+ {{1}\over{6}}(3 X - X^3)(3 Y - Y^3) \rho^3 \nonumber \\
&&\phantom{{{1}\over{2 \pi}} e^{-\haf X^2}e^{-\haf Y^2}  1  }
+ {{1}\over{24}}(3 - 6 X^2 + X^4)(3 - 6 Y^2 + Y^4) \rho^4 
... \Big] . \nonumber 
\label{bivariate_Gaussian_dist_expansion}
\end{eqnarray}
Note that the coefficient of each term in this expansion over $\rho$
can be separated into two factors that depend 
on either $X$ alone or $Y$ alone.
The extension to higher powers of $\rho$ is straightforward,
and the higher-order coefficients have this property as well.

As noted above, I assume that 
the characteristic curve is antisymmetric: $\hat X(-X)=-\hat X(X)$.
The integral $\Upsilon_{XY}$ (Eq.\ \ref{r_M_exact}) involves first powers of the
functions $\hat X(X)$ and $\hat Y(Y)$.
In this integral,
only terms odd in both $X$ and $Y$ match the antisymmetry of the characteristic
curve, and yield a nonzero result.
Such terms are also odd in $\rho$,
as is seen from inspection of Eq.\ \ref{PDistExpansion}.
The first-order terms thus involve the
integrals:
\begin{eqnarray}
B_X &\equiv & \int dX {{1}\over{\sqrt{2\pi}}} e^{-\haf X^2} \hat X(X) X \label{B_def} \\
B_Y &\equiv & \int dY {{1}\over{\sqrt{2\pi}}} e^{-\haf Y^2} \hat Y(Y) Y , \nonumber
\end{eqnarray}
where again $X$ and $Y$ 
can stand for either real or imaginary parts of $x$ and $y$.
Here, I consider terms up to order 3 in $\rho$.
One thus encounters the further integrals:
\begin{eqnarray}
D_X &\equiv & \int dX {{1}\over{\sqrt{2\pi}}} e^{-\haf X^2} \hat X(X) X^3 , \label{D_def} 
\end{eqnarray}
and the analogous expression for $D_Y$.
Therefore, through fourth order in $\rho$:
\begin{eqnarray}
\langle \hat r_M\rangle 
=\Upsilon_{XY}
&\approx & \int dX\;dY\; {{1}\over{2\pi}}
e^{-\haf X^2}
e^{-\haf Y^2} \hat X(X) \hat Y(Y) \label{hat_rM_bar_approx} \\
&&\times\left[(XY) \rho +{{1}\over{6}}(3 X-X^3)(3 Y-Y^3) \rho^3
+... \right] \nonumber \\
&\approx &B_X B_Y \,\rho + {{1}\over{6}} (3 B_X-D_X)(3 B_Y-D_Y) \,\rho^3 + ... . \nonumber 
\end{eqnarray}
For thresholds $v_o\approx 1$ the 
linear approximation is quite accurate \citep{coo70,tms86,jen98};
however, for other values of $v_o$ the higher-order terms 
can become important.
Our notation differs from that of previous
authors;
our $B_X B_Y$ is equal to $\left[(n-1)E+1\right]$ of \citet{coo70} and \citet{tms86}.
It is equal to the $A{{\hat \sigma^2}\over{\sigma^2}}$ of \citet{jen98}.

Figure\ \ref{re_plot} shows typical results of
the expansion of Eq.\ \ref{hat_rM_bar_approx},
for a 4-level correlator,
and compares this estimate for $\hat r_4(\rho)$ with the results from direct
integration of Eq.\ \ref{r_M_exact}
over rectangles in the $x-y$ plane.
In this example, $v_{0x}=v_{0y}\equiv v_0$.
For both $v_0=1$ and $v_0=0.602$,
$\langle \hat r_M\rangle$ is relatively flat,
with a sharp upturn very close to $\rho=1$.
However, in both cases, but especially for $v_0=0.602$,
the curve of $\hat r_4$ bends upward well before $\rho=1$,
so that the linear approximation is good only for relatively small $\rho$.

\subsubsection{Noise: Approximate Results}\label{noise_approx_sec}

Expressions for the noise in the integral
involve the integral $\Upsilon_{X2Y2}$ (Eq.\ \ref{sds_re_im_r_M}).
This integral involves only the squares of the 
characteristic curves $\hat X(X)^2$ and $\hat Y(Y)^2$.
Because the characteristic curves are antisymmetric about 0,
their squares are symmetric: $\hat X(X)^2=\hat X(-X)^2$ and $\hat Y(Y)^2=\hat Y(-Y)^2$.
Therefore,
the only contributions come from 
terms in the expansion of $P(x,y)$
(Eq.\ \ref{PDistExpansion})
that are even in $X$ and $Y$.
One thus encounters the integrals:
\begin{eqnarray}
A_{X2} &\equiv &\int dX\; {{1}\over{\sqrt{2 \pi}}}e^{-\haf X^2}\; \hat X(X)^2 \label{ACE_def} \\
C_{X2} &\equiv &\int dX\; {{1}\over{\sqrt{2 \pi}}}e^{-\haf X^2}\; \hat X(X)^2 X^2 ,  \nonumber \\
E_{X2} &\equiv &\int dX\; {{1}\over{\sqrt{2 \pi}}}e^{-\haf X^2}\; \hat X(X)^2 X^4 ,  \nonumber
\end{eqnarray}
and analogously for $A_{Y2}$, $C_{Y2}$, and $E_{Y2}$.
Then, through fourth order in $\rho$,
\begin{eqnarray}
\Upsilon_{X2Y2}
&=&\int dX\;dY\;P(X,Y)\; \hat X(X)^2\, \hat Y(Y)^2 \label{Upsilon_X2Y2_approx} \\
&\approx& \int dX\;dY\;{{1}\over{2 \pi}}
e^{-\haf X^2}
e^{-\haf Y^2} \hat X(X)^2\, \hat Y(Y)^2 \nonumber \\
&&\quad \times\left[1 + \haf (1-X^2)(1-Y^2)\rho^2 +
\otf (3-6X^2+X^4)(3-6Y^2+Y^4)\rho^4
...\right] \nonumber \\
&\approx& A_{X2}A_{Y2}+\haf(A_{X2}-C_{X2})(A_{Y2}-C_{Y2})\rho^2+\otf(3 A_{X2}-6 C_{X2}+E_{X2})(3 A_{Y2}-6 C_{Y2}+E_{Y2})\rho^4... . \nonumber
\end{eqnarray}
%
I find
the standard deviations of real and imaginary parts from 
Eqs.\ \ref{sds_re_im_r_M},\ \ref{hat_rM_bar_approx}, and\ \ref{Upsilon_X2Y2_approx}:
\begin{eqnarray}
\langle\re[\hat r_M]^2\rangle-\langle\re[\hat r_M]\rangle^2 &\approx&
{{1}\over{2 N_q}} \Bigl(\Bigl\{ A_{X2}A_{Y2} \Bigr\} \label{hat_rM_sd_approx}\\
&&\phantom{{{1}\over{2 N_q}} \Bigl( }
+\Bigl\{\haf(A_{X2}-C_{X2})(A_{Y2}-C_{Y2})-B_{X}^2B_{Y}^2\Bigr\} \rho^2 \nonumber \\
&&\phantom{{{1}\over{2 N_q}} \Bigl( }
+\Bigl\{\otf(3 A_{X2}-6 C_{X2}+E_{X2})(3 A_{Y2}-6 C_{Y2}+E_{Y2}) \nonumber \\
&&\phantom{{{1}\over{2 N_q}} \Bigl(+\Bigl\{ } \quad
-\thd B_X(3 B_X-D_X) B_Y(3 B_Y-D_Y)\Bigr\}\rho^4
...
\Bigr) \nonumber \\
\langle\im[\hat r_M]^2\rangle
&\approx &{{1}\over{2 N_q}}\Bigl( \Bigl\{ A_{X2}A_{Y2} \Bigr\} \nonumber \\
&&\phantom{ {{1}\over{2 N_q}}\Bigl( } -\Bigl\{ B_X^2B_Y^2 \Bigr\} \rho^2 \nonumber \\
&&\phantom{ {{1}\over{2 N_q}}\Bigl( } -\Bigl\{\thd B_X(3 B_X-D_X) B_Y(3 B_Y-D_Y)\Bigr\}\rho^4
...
\Bigr) . \nonumber
\end{eqnarray}
I have used the fact that 
$\langle \hat r_M\rangle$ is purely real;
this is a consequence of the assumption that $\rho$ is purely real.

\begin{figure}[t]
\plotone{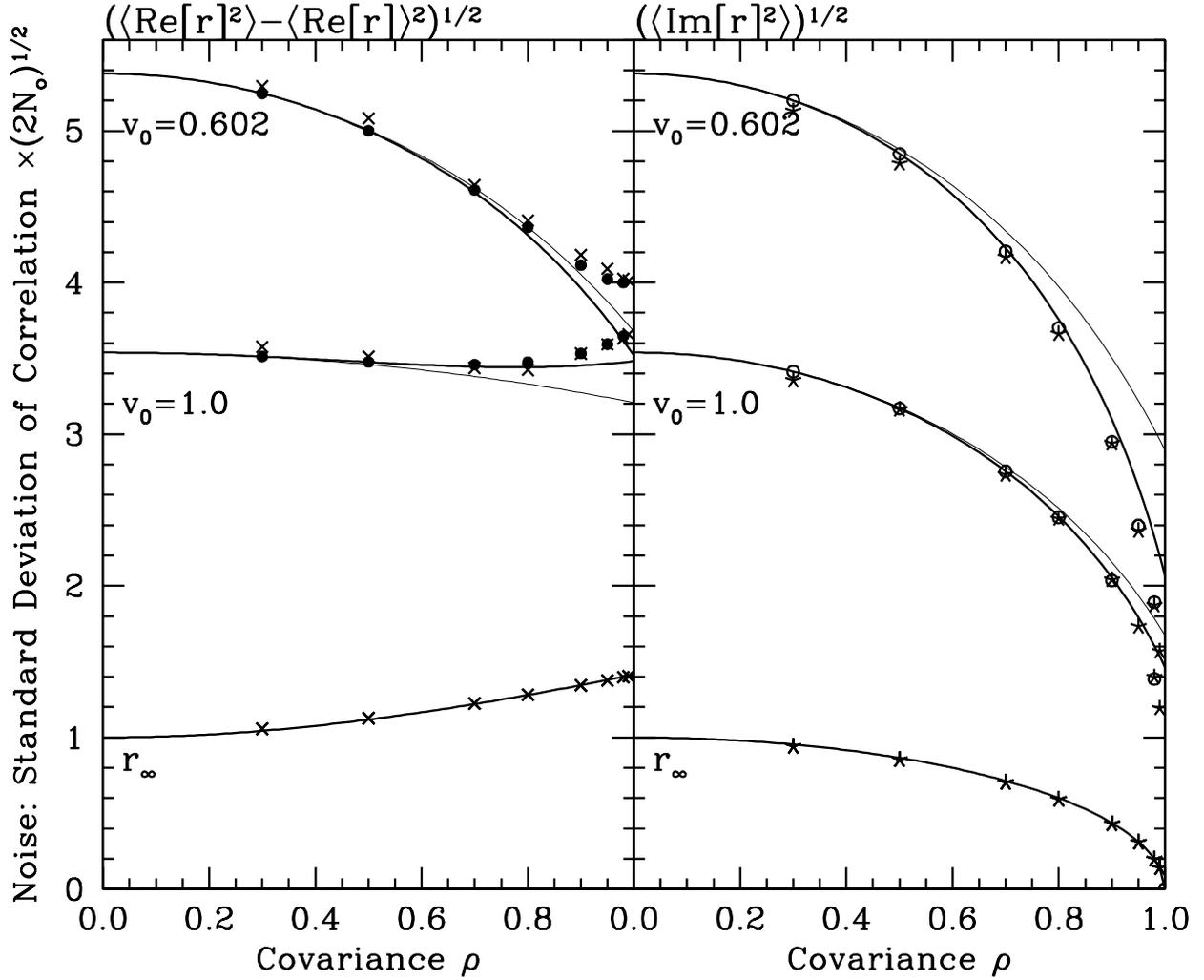}
\figcaption[]{
Standard deviation of the real part of correlation (left panel),
and of the imaginary part (right panel), plotted with covariance $\rho$.
Both are normalized for the number of samples
by the factor $\sqrt{2 N_q}$.
Curves show $\langle\hat r_{\infty}\rangle$,
and approximations to
$\langle\hat r_4\rangle$ 
for quantization with $v_0=1.0$ and $v_0=0.602$,
with $n=3$.
Heavy lines show the fourth-order approximation of Eq.\ \ref{hat_rM_sd_approx},
and light lines show the approximation to second order.
The values found by \citet{coo70} are the y-intercepts ($\rho=0$).
Filled circles show true values as computed by direct integration of Eq.\ \ref{sds_re_im_r_M}.
Crosses show results of simulations (\S \ref{simulations}).
\label{sd_plot}}
\end{figure}

Figure\ \ref{sd_plot} shows examples 
of the standard deviations of $\re [\hat r_{4}]$
and $\im [\hat r_{4}]$ for 2 choices of $v_0$.
These are the noise in estimates of the correlation.
Note that the noise varies with $\rho$.
The quadratic variation of these quantities with $\rho$ is readily apparent.
The higher-order variation is more subtle, although it does lead to an
upturn of the standard deviation of $\re [\hat r_{4}]$ near $\rho\approx 0.7$,
for $v_0=1$.
The series expansions become inaccurate near $\rho=1$, as expected.
The standard deviation of 
$\re [\hat r_{4}]$ 
can also increase, instead of decrease, for large $\rho$.
Such an increase is more common for parameter choices with $v_0>1$.
Again, note that
the standard deviation of the imaginary part always decreases
with increasing $\rho$.

\subsubsection{SNR for Quantized Correlation}

The signal-to-noise ratio (SNR) for a
quantizing correlator is 
the quotient of 
the mean and variance of $\hat r_M$:
the results
of \S\ \ref{mean_approx_sec} and\ \ref{noise_approx_sec}.
I recover the results of
\citet{coo70} for the SNR
for a quantizing correlator, by using our approximate expressions
through first order in $\rho$
(see also \citet{hag73}, \citet{tms86}, \citet{jen98}):
\begin{equation}
{\cal R}_{M}(\re [\hat r_M]) \approx {{\langle \re [\hat r_M] \rangle}\over{\sqrt{\langle \re [\hat r_M]^2\rangle- \langle \re [\hat r_M]\rangle^2}}} 
\approx \sqrt{2 N_q}{{B_X B_Y}\over{\sqrt{A_{X2}A_{Y2}}}} \rho.
\end{equation}
For a 4-level correlator,
a SNR of ${\cal R}_{4}=0.88115 \times \rho$ is attained for $n=3$, $v_0$=1,
in the limit $\rho\rightarrow 0$.
Many 4-level correlators use these values.
The maximum value for ${\cal R}_{4}$ is actually obtained
for $n=3.3359$, $v_0=0.9815$, for which ${\cal R}_{4}=0.88252 \times \rho$.
This adjustment of quantization constants provides a 
very minor improvement in SNR.

\begin{figure}[t]
\plotone{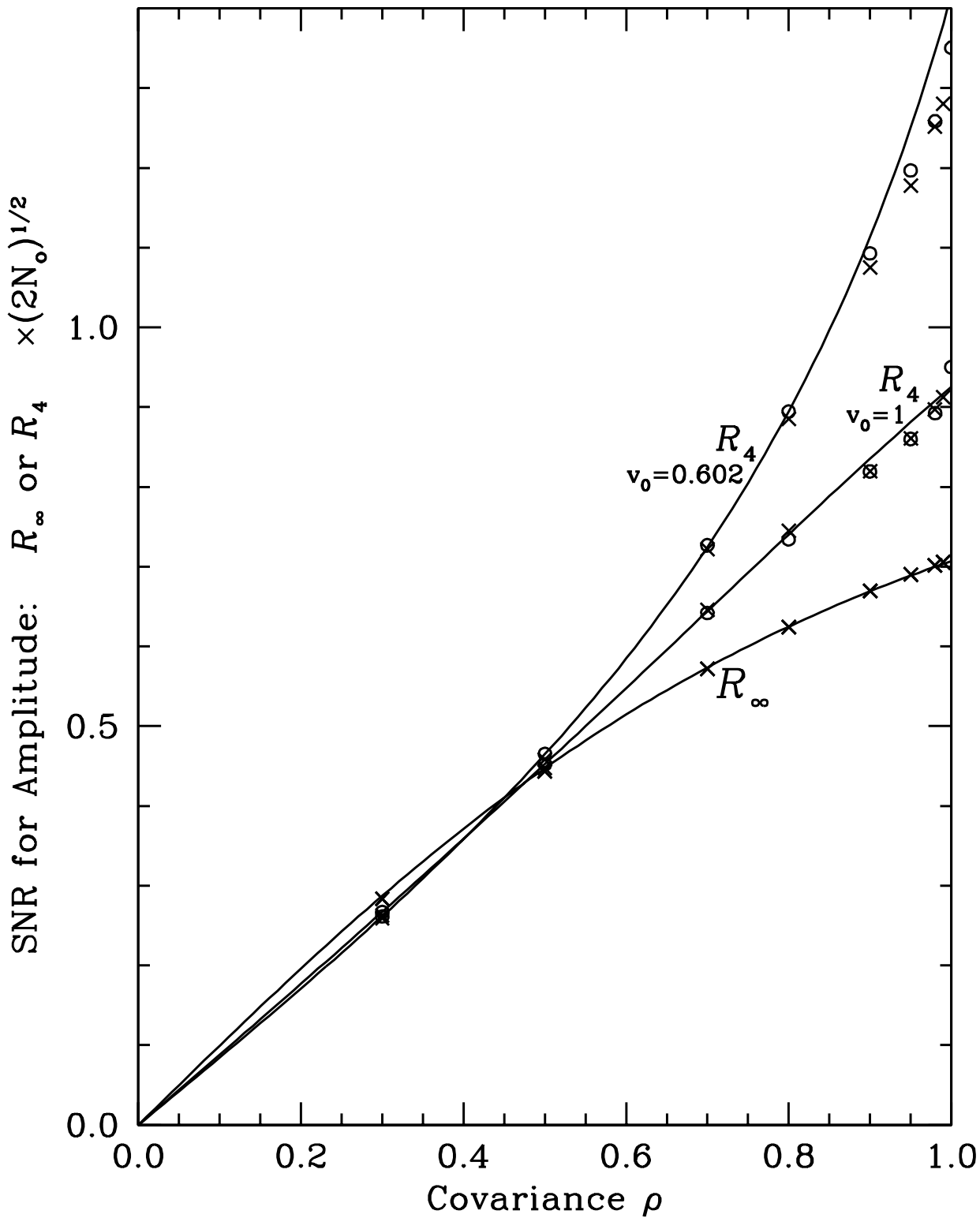}
\figcaption[]{
Signal-to-noise ratio ${\cal R}_{\infty}$ or ${\cal R}_{4}$
normalized for number of samples by $1/\sqrt{2 N_q}$ and
plotted with covariance $\rho$.
Curves mark the exact expression given 
by Eq.\ \ref{SNRinfty} for ${\cal R}_{\infty}$,
or the approximate expression 
given by the expansions of Eqs.\ \ref{hat_rM_bar_approx} 
and\ \ref{hat_rM_sd_approx} through 
fourth order in Eq.\ \ref{SNR_rM}.
Dots give exact values as found from direct integration of 
Eq.\ \ref{sds_re_im_r_M} in Eq.\ \ref{SNR_rM}.
Crosses show results of computer simulations in \S\ \ref{simulations}.
Quantization uses the characteristic curve in Fig.\ \ref{4_level}
with 
values of $v_0=1.0$ or $v_0=0.602$, and $n=3$.
\label{SNR_plot}}
\end{figure}

For nonvanishing $\rho$,
the optimum level settings depend upon the covariance $\rho$.
The signal-to-noise ratio is 
\begin{eqnarray}
{\cal R}_{M}(\re [\hat r_M])&=&
\sqrt{2 N_q} {{\Upsilon_{XY}}\over{\sqrt{\Upsilon_{X2Y2}-(\Upsilon_{XY})^2}}} .
\label{SNR_rM}
\end{eqnarray}
This can be approximated using the expansions for $\Upsilon_{XY}$ and $\Upsilon_{X2Y2}$;
Figure\ \ref{SNR_plot} shows the results.
Note that in the examples in the figure, the SNR for the quantized correlations 
actually curve above that for the unquantized correlation beyond $\rho\approx 0.5$;
this indicates that correlation of quantized signals can actually
yield higher signal-to-noise ratio than would be obtained from correlating the 
same signals before quantization.
This results from the decline in noise with increasing $\rho$
visible in Fig.\ \ref{sd_plot}.

For a proper comparison of SNRs,
one must compare with the SNR obtained for non-quantized correlation,
${\cal R}_{\infty}(\re [\hat r_M])$, Eq.\ \ref{SNRinfty}.
One finds:
\begin{eqnarray}
{{ {\cal R}_{M}}(\re [\hat r_M])\over{ {{\cal R}_{\infty}(\re [\hat r_M])}}}
={{\Upsilon_{XY}}\over{\sqrt{\Upsilon_{X2Y2}-(\Upsilon_{XY})^2}}}
{{\sqrt{1-\rho^2}}\over{\rho}} .
\end{eqnarray}
Figure\ \ref{ratio_plot}
shows this ratio for 2 choices of $v_0$.
The ratio can exceed 1, again indicating that quantized correlation 
provides a more accurate result than would correlation of
an unquantized signal.

The SNR for a measurement of phase for quantized correlation
is the inverse of the standard deviation of the phase,
as discussed in \S\ \ref{SNRinfty}.
For quantized correlation, this is
\begin{equation}
{\cal R}_{M}(\phi [r_{M}]) = 
{{\re [r_{M}]}\over{\sqrt{\langle \im [r_{M}]^2\rangle}}}
={\sqrt{2 N_q}} {{\Upsilon_{XY}}\over{\sqrt{A_{X}A_{Y}-(\Upsilon_{XY})^2}}} .
\end{equation}
Again, because $\Upsilon_{XY}$ always increases with $\rho$,
the SNR of the phase always increases with increasing covariance.

The ratio of the SNR for phase to that for correlation
of an unquantized signal, 
${ {\cal R}_{M}}(\phi [\hat r_M])/{ {{\cal R}_{\infty}(\phi [r_{\infty}])}}$
(see Eq.\ \ref{SNRphase}),
provides an interesting comparison.
For $\rho\rightarrow 0$, the statistics for the imaginary part of
the correlation are identical to those for the real part (as they must be),
and the highest SNR for the phase is given by the 
quantizer parameters that are optimal for the real part,
traditionally $v_0=1$, $n=3$.
This ratio is approximately constant with $\rho$ up
to $\rho\approx 0.6$, and then decreases rather rapidly.
Simulations suggest that quantized correlation is less
efficient than unquantized for measuring phase;
however I have not proved this in general.

\section{SIMULATIONS}\label{simulations}

Simulation of a 4-level correlator provides a useful perspective.
I simulated such a correlator by 
generating two sequences of random, complex numbers,
$x_i$ and $y_i$.
The real parts of $x_i$ and $y_i$
are drawn from one bivariate Gaussian distribution,
and their imaginary parts from another independent one 
(see Eq.\ \ref{product_2_bivariates}).
These bivariate Gaussian distributions can be described equivalently
as elliptical Gaussian distributions,
with major and minor axes inclined 
to the coordinate axes $X=\re[x_i]$ and $Y=\re[y_i]$
and the corresponding axes 
for the imaginary parts.

\citet{mey75} gives expressions 
that relate the semimajor and semiminor axes
and angle of inclination
of an elliptical Gaussian distribution
to the 
normalized covariance $\rho$
and variances $\sigma_X^2$ and $\sigma_Y^2$.
For the special case of $\sigma_X=\sigma_Y=1$
used in this work,
the major axis always lies at angle $\pi/4$ to both coordinate axes,
along the line $X=Y$.
The semimajor axis $b_1$ and semiminor axis $b_2$ are then given by
\begin{eqnarray}
b_1 &=& \sqrt{1+\rho}, \label{rotated_ellipse} \\
b_2 &=& \sqrt{1-\rho}. \nonumber
\end{eqnarray}

To form the required elliptical distributions,
I drew pairs of elements from a circular Gaussian distribution,
using the Box-Muller method
(see \citet{pre89}).
I scaled these random elements so that their standard deviations
were $b_1$ and $b_2$.
I then rotated the resulting 2-element vector by $\pi/4$
to express the results in terms of $X$ and $Y$.
I repeated the procedure for the imaginary part.

I quantized the sequences $x_i$ and $y_i$ 
according to the 4-level characteristic curve
shown in Figure\ \ref{4_level}, to yield $\hat x_i$ and $\hat y_i$.
Both the unquantized and the quantized sequences were correlated
by forming the products $x_i y_i^*$ and $\hat x_i \hat y_i^*$,
respectively,
and results were averaged 
over $N_q=10^5$ instances of the index $i$.
This procedure yields one realization each of
$r$ and $\hat r_4$. I found that values
for $N_q$ smaller than about $10^5$ 
could produce significant departures
from Gaussian statistics for $\hat r_4$,
particularly for
larger values of $\rho$.

I repeated the process to obtain 4096 
different realizations of $r$ and $\hat r_4$.
I found the averages and standard deviations
for the real and imaginary parts for this set of realizations.
Figures\ \ref{re_plot} through\ \ref{ratio_plot} show 
these statistical results of the simulations,
and compare them with the mathematical results of
the preceding sections.  
Clearly, the agreement is good.

In graphical form,
samples of the correlation form an elliptical Gaussian
distribution in the complex plane,
centered at the mean value of correlation $\langle r_{\infty}\rangle$
or $\langle \hat r_{M}\rangle$,
as the case may be.
The principal axes of the distribution lie along
the real and imaginary directions
(or, more generally, the directions in phase with $\rho$ and out of
phase with $\rho$).
The lengths of these principal axes are the variances
of real and imaginary parts.

\section{DISCUSSION AND SUMMARY}

\subsection{Change of Noise with Covariance}

The fundamental result of this paper is that a change in covariance $\rho$
affects quantized correlation $\hat r_M$
differently from unquantized
correlation $r_{\infty}$.  
For unquantized correlation, an increase in covariance
$\rho$ increases noise for estimation of signal amplitude.  For
quantized correlation, an increase in $\rho$ can increase or decrease
amplitude noise.  For both quantized and unquantized
correlation, an increase in $\rho$ leads to a decrease in phase noise.  
In this work, I arbitrarily set the phase of $\rho$ to 0, so
that amplitude corresponds to the real part, and phase to the
imaginary part of the estimated correlation.  The net noise (summed,
squared standard deviations of real and imaginary parts) can decrease
(or increase) with increasing $\rho$: noise is not conserved.

I present expressions for the noise as a function of quantization
parameters, both as exact expressions that depend on $\rho$ and on
power-series expansions in $\rho$.  These expressions, and a
power-series expansion for the mean correlation, are given through
fourth order in $\rho$.

The increase in noise with covariance $\rho$ for analog correlation,
sometimes called source noise or self-noise, is sometimes ascribed to
the contribution of the original, noiselike signal to the noise of
correlation.  This idea is difficult to generalize to comparisons
among quantized signals, because such comparisons require additional
assumptions about changes in quantizer levels and the magnitude of the
quantized signal when the covariance changes.  These comparisons are
simpler for multiple correlations derived from a single signal (as,
for example, for the correlation function of a spectrally-varying
signal), and I will discuss them in that context elsewhere
\citep{gwi03}.  The discussion in this paper is limited to ``white''
signals, without spectral variation and with only a single independent
covariance.

\subsubsection{Increase in SNR via Quantization}

One interesting consequence of these results 
is that signal-to-noise ratio of correlation can actually
be greater for quantized signals, than it would be for
correlation of the same signals before quantization.  
At small covariance $\rho$, SNR is always lower for quantized signals, but
this need not be the case for covariance $\rho \gtwid 0.4$.
This appears to
present a paradox, because the process of quantization intrinsically
destroys information: the quantized signals $\hat x_i$, $\hat y_i$
contain less information
than did the original signals $x_i$, $y_i$.
However, 
correlation of unquantized signals also destroys information:
it converts $x_i$ and $y_i$ to the single quantity $c_i=(x_i y_i^*)$.
Different information is destroyed in the two cases.

Moreover, correlation does not always yield the most accurate estimate
of the covariance $\rho$.
As a simple example, consider the series $\{X_i\}=\{0.400,-0.800,1.600\}$
and $\{Y_i\}=\{0.401,-0.799,1.600\}$.
Here, $N_q=3$.
One easily sees that $X$ and $Y$ are highly correlated.
If $X$ and $Y$ 
are known to be drawn from Gaussian distributions with unit standard deviation,
Eq.\ \ref{rotated_ellipse}
suggests that $\rho\approx 0.999$.
However, the correlation is $r={{1}\over{3}}\sum_i X_i Y_i = 1.29$.
Clearly $r$ is not an optimal measurement of $\rho$.
I will discuss strategies for optimal estimates of covariance elsewhere \citep{gwi03}.

\subsubsection{Quantization Noise}

Sometimes effects of quantization are described as 
``quantization noise'': an additional source of noise
that (like ``sky noise'' or ``receiver noise'') reduces
the correlation of the desired signal.
However, unlike other sources of noise,
quantization destroys information in the signals,
rather than adding unwanted information.
The discussion of the preceding section suggests that
the amount of information that quantization destroys
(or, more loosely, the noise that it adds)
depends on what information is desired;
and that correlation removes information as well.
Unless the covariance $\rho$ is small,
effects of quantization cannot be
represented as a one additional, independent source of noise,
in general.

\subsubsection{Applications}

The primary result of this paper is that 
for quantized correlation, noise can increase or
decrease when covariance increases; whereas for continuous signals
it increases.  This fact is important for applications
requiring accurate knowledge of the noise level; 
as for example in studies of
rapidly-varying strong sources such as pulsars, where one wishes to
know whether a change in correlation 
represents a significant change in the pulsar's emission;
or for single-dish or interferometric
observations of intra-day variable sources,
where one wishes to know whether features that may appear and disappear 
are statistically
significant.

A second result of this paper is that the signal-to-noise ratio for quantized
correlation can be quite different from that expected for a continuum
source, or for a continuum source with added noise.  This effect is
most important for large correlation, $\rho\gtwid 0.5$.  Correlation
this large is often observed for strong sources, such as 
the strongest
pulsars, or maser lines; and for the strongest continuum sources
observed with sensitive antennas.  For example, at Arecibo Observatory
a strong continuum source easily dominates the system temperature, at
many observing frequencies.  The effect
will be even more common for some proposed designs of the 
Square Kilometer Array.

Many sources show high correlation that varies dramatically with frequency;
such sources include scintillating pulsars, and maser lines.
Typically observations of these source involve determination
of the full correlation function, and a Fourier transform
to obtain an estimated cross-power or autocorrelation spectrum.
I discuss the properties of noise for this analysis
elsewhere.

\subsubsection{SNR Enhancement?}

An interesting question is whether one can take advantage of the
higher SNR afforded by quantization at high $\rho$ even for
weakly-correlated signals, perhaps by adding an identical signal
to each of 2 weakly covariant signals and so
increasing their covariance $\rho$, before quantizing them.  
The answer appears to be ``no''.
As a simple example, consider a pair of signals with covariance
of $\rho\approx 0.01$.  After correlation of 
$N_q=2\times 10^6$ instances of the signal, 
using a 4-level correlator with $v_0=1$ and $n=3$,
the SNR is 20,
and one can determine at a level of 2 standard deviations whether
$\rho=0.010$ or $\rho=0.011$.  If a single signal, with 4.6 times
greater amplitude, is added to both of the original signals, then these 2 cases
correspond to $\rho=0.7004$ or $\rho=0.7005$. 
To distinguish them at 
2 standard deviations requires a SNR of 1400, requiring
$N_q=4\times 10^8$ samples of the quantized correlation.
Thus, the increase in SNR is more than outweighed by the reduction
in the influence on the observable.

\subsection{Summary}

In this paper, I consider 
the result of quantizing and correlating two complex noiselike signals,
$x$ and $y$
with normalized covariance $\rho$.
The signals are assumed to be statistically stationary, ``white,''
and sampled at the Nyquist rate.
The correlation $r$ 
provides a measurement of $\rho$.
The variation of $r$ about that mean,
characterized by its standard deviation,
provides a measure of the random part of the measurement,
or noise.

I suppose that the signals $x$ and $y$ are quantized 
to form $\hat x$ and $\hat y$.
I suppose that the characteristic curves that govern
quantization are antisymmetric, 
with real and imaginary parts subject to the same
characteristic curve.  I recover the classic
results for the noise 
for $\rho =0$, and for the mean correlation, to first order in
$\rho$, in the limit $\rho\rightarrow 0$
\citep{van66,coo70,hag73,tms86}.  
I find exact expressions
for the mean correlation and the noise, and approximations valid through
fourth order in $\rho$.
I compare results with simulations.  Agreement is excellent for the
exact forms, and good for $\rho$ not too close to 1, for the
approximate expressions.

I find that for nonzero values of $\rho$,
the noise varies, initially quadratically, with $\rho$.
I find that the noise in an estimate of the amplitude of $\rho$ 
can decrease with increasing $\rho$;
this is opposite the behavior of noise for correlation of unquantized signals,
for which noise always increases with $\rho$.
The mean correlation can increase more rapidly than linearly with $\rho$.
The signal-to-noise ratio (SNR) 
for correlation of quantized signals can be greater than that for 
correlation of unquantized signals, for $\rho\gtwid 0.5$.
In other words, correlation of quantized signals can be more
efficient than correlation of the same signals before quantization,
as a way of determining the covariance $\rho$.

\acknowledgments

I am grateful 
to the DRAO for supporting this work with extensive
correlator time.
I also gratefully acknowledge the VSOP Project, which is led by the Japanese 
Institute of Space and Astronautical Science in cooperation with many 
organizations and radio telescopes around the world.
I thank an anonymous referee for useful comments.
The U.S. National Science Foundation provided financial support.

\newpage
\figurenum{1}

\figurenum{2}

\figurenum{3}

\figurenum{4}

\figurenum{5}

\end{document}